\documentclass[conference]{IEEEtran}
\IEEEoverridecommandlockouts

\usepackage[a4paper,
            left=0.509in,
            right=0.509in,
            top=0.75in,
            bottom=1.69in]{geometry}
            
\usepackage{graphicx}
\usepackage{textcomp}
\usepackage{array,ragged2e}
\def\BibTeX{{\rm B\kern-.05em{\sc i\kern-.025em b}\kern-.08em
    T\kern-.1667em\lower.7ex\hbox{E}\kern-.125emX}}

\usepackage{colortbl}
\usepackage{makecell}
\usepackage{pgfplots}
\pgfplotsset{compat=1.17}
\usepackage{amsmath}
\usepackage{multicol}
\usepackage{multirow}
\usepackage{tabto}
\usepackage{wrapfig}
\usepackage{amssymb}
\usepackage{amsmath}
\usepackage{float}
\usepackage{graphicx}
\usepackage[misc,geometry]{ifsym}

\usepackage{textcomp}
\usepackage{array,ragged2e}
\usepackage{pgfplots}
\usepackage{url}
\usepackage{xurl}
\usepackage{lineno,hyperref}
\hypersetup{
  linkcolor  = red!60!black,
  citecolor  = blue!90!white,
  urlcolor   = violet!60!black,
  colorlinks = true,
}

\usepackage{algorithm}
\usepackage{algpseudocode}

\usepackage{hyperref}
\usepackage{svg}
\usepackage{xcolor}
\newcommand{\orcid}[1]{\href{https://orcid.org/#1}{\includesvg[width=10pt]{orcid.svg}}}

\usepackage{threeparttable}

\usepackage{pgf-pie}  

\usepackage{float}
\newcommand{\tsuper}{\textsuperscript}
\usepackage{enumitem}
\newtheorem{rule2}{Rule}
\usepackage{orcidlink}

\begin{document}
\title{Graph-Based Optimisation of Network Expansion in a Dockless Bike Sharing System}
\author{

    \IEEEauthorblockN{Mark Roantree\orcidlink{0000-0002-1329-2570}}
    \IEEEauthorblockA{Insight Centre for Data Analytics \\School of Computing, Dublin City University \\Dublin, Ireland \\
    Email: mark.roantree@dcu.ie}
\\
    \IEEEauthorblockN{Dinh Viet Cuong}
    \IEEEauthorblockA{Insight Centre for Data Analytics \\School of Computing, Dublin City University \\Dublin, Ireland \\
    Email: dinh.cuong2@mail.dcu.ie}
\and
    \IEEEauthorblockN{Niamh Murphy}
    \IEEEauthorblockA{School of Computing \\Dublin City University, Dublin, Ireland \\
    Email:niamh.murphy88@mail.dcu.ie }
\\
\\
    \IEEEauthorblockN{Vuong M. Ngo\orcidlink{0000-0002-8793-0504}~\textsuperscript{\Letter}}
    \IEEEauthorblockA{Ho Chi Minh City Open University \\Ho Chi Minh City, Vietnam\\
    Email: vuong.nm@ou.edu.vn}
}

%
%
%
%
\maketitle              
\begin{abstract}
Bike-sharing systems (BSSs) are deployed in over a thousand cities worldwide and play an important role in many urban transportation systems. BSSs alleviate congestion, reduce pollution and promote physical exercise. It is essential to explore the spatiotemporal patterns of bike-sharing demand, as well as the factors that influence these patterns, in order to optimise system operational efficiency. In this study, an optimised geo-temporal graph is constructed using trip data from Moby Bikes, a dockless BSS operator. The process of optimising the graph unveiled prime locations for erecting new stations during future expansions of the BSS. The Louvain algorithm, a community detection technique, is employed to uncover usage patterns at different levels of temporal granularity. The community detection results reveal largely self-contained sub-networks that exhibit similar usage patterns at their respective levels of temporal granularity. Overall, this study reinforces that BSSs are intrinsically spatiotemporal systems, with community presence driven by spatiotemporal dynamics. These findings may aid operators in improving redistribution efficiency. 
\end{abstract}

\begin{IEEEkeywords}
Bike Sharing, Urban Movement, Graph Theory, Clustering, Community Detection, spatiotemporal Analysis
\end{IEEEkeywords}

%
%
\section{Introduction}

Commercially viable technology capable of tracking the movement of bikes has enabled the rapid expansion of Bike Sharing Systems (BSS) in cities across many countries \cite{LIN2018258}. By August 2022, the number of bikes in these schemes had grown to almost 9 million across 1,914 systems, spanning 92 countries and 1,590 cities \cite{BSWM:2022}. This substantial growth has transformed the bike sharing market into a high performing industry, which according to Statista \cite{BikeSharing2024} has a valuation of 9.46 billion US dollars. However, as bike networks grow, improving their efficiency to meet the demands of users is crucial for the survival of BSS operators \cite{art1}. One of the primary objectives of this endeavor is to identify the optimal number of stations in suitable locations, a task that necessitates spatial analysis employing appropriate tools and methods. Graph networks have been widely utilized in various applications, including textile structure \cite{Helmer-Ngo:2015}, \cite{Ngo_2021a}, knowledge-based systems \cite{Nguyen_2014}, \cite{Ngo_2021b}, and electronic health records \cite{Ngo_2024}. In the context of bike-sharing systems, graph-based approaches have been employed to model locations, trips, and associated time intervals, facilitating the investigation of system dynamics. In these projects, bike journeys were modelled as network structures and through the analysis of these networks, the inherent characteristics of journey dynamics were studied \cite{Lei, Ghandeharioun, Seo, Yang2019}. We examine these dynamics by representing spatial locations as nodes, with trips forming the edges between starting and ending locations.

\subsection{Problem Statement}
An in-depth understanding of usage patterns can greatly improve the impact of management strategies, making the service more attractive to users. However, two key obstacles exist: the availability of data with the necessary features and the ability to model the complexities of the network dynamics \cite{f1}. The first problem requires either some form of online data acquisition such as \cite{MMCR19} where live data is processed into temporally aware data marts, or, collaboration with the bike rental company. The second problem requires a comprehensive spatiotemporal characterisation of the travel patterns in order to identify high usage locations with no station on the network. 

The data issue was resolved by collaborating with Moby Bikes\footnote{\url{https://mobybikes.com/}} who, after anonymisation, made all data available for the purpose of this research. Trips made using Moby’s fleet generated significant quantities of time and location specific data to enable the study of travel behaviour and mobility at trip level granularity. The multi-dimensional, interconnected nature of the spatiotemporal data motivated the use of a graph database to resolve the second issue. The  Neo4j \cite{Neo4j2023} database was used to construct a number of different graph networks and to facilitate graph based (community detection) clustering as a means of understanding network traffic and to ensure that newly created stations observed the same activity patterns as existing stations. Given the data and infrastructural solution available, the research questions are: can trip data be modelled so as to identify optimal candidate locations for network expansion (new stations)? can spatial analysis be facilitated at different levels  of temporal granularity? and finally, is there a means of validating new stations so that they are not outlier stations with activity patterns unrepresentative of other nodes in the network (outliers)?

\subsection{Contribution and Paper Structure}
The Moby network comprises 92 fixed stations and in prior work \cite{CNCR2024}, a distance function (discussed later as eq. \ref{eqn:mod_eq}) was employed to reassign non-station source and destination nodes to their closest fixed station.  This enabled the construction of spatiotemporal graphs to facilitate time-based comparison of different stations and the construction of station profiles to model their interactions with all other stations. However, it could not identify precise locations for the placement of new fixed stations to reduce traffic bottlenecks. Analysis showed that many virtual stations could be created to accommodate the lack of fixed stations in key locations but this represents uncontrolled expansion of the network. The contribution of this work can be articulated as follows:
\begin{itemize}
    \item Optimizing the expansion of the network through a novel algorithm for identifying new station locations;
    \item A methodology for facilitating spatiotemporal analysis using graphs with different temporal granularities; 
    \item A validation mechanism using community detection to ensure that new stations had similar behaviour or characteristics as existing stations.
\end{itemize}

\textbf{Paper Structure.} 
The remainder of this paper is organised as follows. Section 2 briefly reviews
relevant studies of BSSs with a spatial, temporal, or graph theory focus. A detailed description of the dataset and the preprocessing steps performed are presented in Section 3. Section 4 describes the main methodologies. A summary of the results and a validation of the results are contained in Section 5. Finally, Section 6 concludes the paper with a discussion and recommendations for future work.

%
%
\section{Related Work}

In recent years there has been a surge in research interest in the spatiotemporal characteristics of BSSs. Sun et al. \cite{sun} studied a number of factors that affected BSS usage and found that BSSs' usage increased during peak hours on weekdays, indicating that BSSs are predominantly used for commuting purposes. Zhou \cite{chic} uncovered a similar finding by constructing a bike flow similarity graph of a Chicago BSS and used a fast greedy algorithm to detect spatial communities with unique travel patterns on weekdays and weekends. Munoz-Mendez et al. \cite{f1} developed a novel modified Infomap clustering approach to capture the spatiotemporal patterns observed in a London-based BSS. Through clustering, self-contained and interconnected community structures were discovered, with approximately 75\% of the observed trips starting and ending within the same community \cite{f1}.  In separate work, the authors used global metrics to capture the overall structure of the network while local metrics were used to identify prominent nodes across the network. Commonly used metrics encompass the count of nodes, edges, degree, and strength, which signify the level of activity and connectivity within a given location \cite{Zhang}, \cite{Yao}.

The previously discussed research focuses on conventional dock-based BSSs. The usefulness of these studies may be limited as users of dockless BSSs are not confined to fixed stations, meaning the spatial distribution is drastically different. There are fewer relevant studies that have analysed the spatiotemporal characteristics of dockless BSSs on real-world trip data. In \cite{Lin_He_Peeta_2018}, Lin et al. constructed a spatially embedded network using dockless BSS data from Beijing. Origin and destination locations were represented by nodes and trip flows by edges. Spatial community detection was used to reduce the complexity of the network. On average, 77\% of trips began and ended inside the same community, indicating that these communities were largely self-contained.

The study conducted by Austwick et al.  \cite{Austwick}  revealed commonalities in the distribution of strength and edge weight within networks generated from diverse bike-sharing systems. In similar work, researchers \cite{Yang2019}, \cite{Jianmin} recommended incorporating a more holistic set of network properties to capture different network features including connectivity metrics such as degree and node flux, spatial distribution using the clustering coefficient algorithm, network stability or connectivity (centrality algorithms), network efficiency, and equity (using the Gini coefficient). More traditional centrality metrics such as \emph{betweenness} and \emph{PageRank}, have also been employed to describe a network's features \cite{Yang2019}.

Limited research has also been conducted using data from Dublin based BSSs. Research has been performed using publicly available data from DublinBikes \cite{JIMENEZ2016228,nuimeprn12013}. These studies classify bike stations according to users’ mobility patterns and analyse usage patterns of stations, respectively.

Previous studies have illustrated the suitability of representing BSS data as a graph. BSSs in cities around the world have been shown to exhibit temporal patterns and largely self-contained sub-networks. However, the majority of the existing literature focuses on dock-based BSS. The few relevant studies that involve dockless BSSs do not group rental locations according to both usage patterns and temporal features. This paper seeks to address this gap in research by analysing a Dublin-based BSS to assess whether observed communities are a result of spatial or temporal features.

Community analysis, a recurrent theme in network research, plays a pivotal role in comprehending a network's structure. Community detection algorithms categorize nodes within the network into distinct communities, where nodes exhibit stronger internal connections than external ones. The Louvain algorithm, as discussed in \cite{Louvain}, stands as the most commonly employed method in this realm. However, Shi et al. \cite{Shi} observed that different algorithms yield diverse community characteristics depending on the measurement criteria. Despite the advancement in network analysis, these methods are predominantly applied to networks with edge connections representing direct interactions. We aim to extend similar methodologies, including visualization and the utilization of complex metrics such as strength, closeness, betweenness, local clustering coefficients, and community detection, to more intricate correlation-based networks.
    
%
%
\section{Data}

    The data used in this research has been collected and provided by Moby Bikes \cite{Moby}. Moby’s bike fleet consists of 95 fully electric bikes, and, with a \emph{dockless} mode or operation, users can collect or drop bikes at any suitable public location. However, to reduce operational overheads, Moby introduced fixed charging stations where customers are financially incentivised to return rented bikes to these fixed locations which we refer to as \emph{stations}. 
    
    The GPS units on the bikes have created a large amount of individual level data which if properly managed, can be modelled as a spatiotemporal graph network. The data collected is stored in two SQL tables, Rental and Location. The Rental table contains information about each logged rental during the period from 3\tsuper{rd} January 2020 to 19\tsuper{th} September 2021. There are 62,324 records in the original Rental table. The Location table provides finer-grained information on each of the locations a bike was either rented from or returned to, during the same 21 month period, with a total of 14,239 records. Interestingly,  COVID-19 was declared a pandemic by the World Health Organisation in March 2020, meaning the majority of the data used was created during the pandemic.

    There were some minor issues with the dataset: references to non-existent data and inadmissible or spurious locations. The dataset was cleaned by removing the following entries:

    \begin{itemize}
        \item Locations outside Dublin (as rentals should only involve journeys within city locations) and rentals that started or ended at these locations.
        \item Locations that are not on land and associated rentals.
        \item Locations that are missing latitude or longitude coordinates and associated rentals.
        \item Rentals that do not report a Rental Location ID or a Return Location ID.
        \item Rentals with a Rental Location ID or a Return Location ID that is not in the Location table.
        \item Location IDs in the Locations table that are not referenced in the Rental table.
    \end{itemize}

\begin{table}[H]
\centering
\caption{Dataset Overview}
\begin{tabular}{|l|r|r|}
\hline
\textbf{Measures} & \textbf{Original Dataset}  & \textbf{Cleaned Dataset}  \\ \hline
Duration of data  & \multicolumn{2}{c|}{Jan 2020 - Sept 2021 ($\sim$21 months)}  \\ \hline
\#stations & 95 & 92     \\ \hline
\#rental & 62,324 & 61,872    \\ \hline
\#location & 14,239 & 14,156  \\ \hline
\end{tabular}
\label{tab:data_summary}
\end{table}

After removing these entries, the Rental table comprised 61,872 entries and the Location table comprised 14,156 entries across 92 stations, as Table \ref{tab:data_summary}. 
    
%
%
    
\section{Methodology}
A methodology was devised as a three-step process: (1) graph construction, involving a preprocessing step to determine the spread of trips and begin/ending locations for generating candidate stations; (2) a process to select suitable stations from the candidate stations, merging them with pre-existing stations. This process is guided by a selection and ranking algorithm; and (3) community detection to understand the impact of the new stations on the network.


\subsection{Graph Construction}
\label{subSec:GraphConstruction}

The majority of bike-sharing systems operate on a dock or station-based model, where customers are required to rent and return bikes at designated stations. Consequently, it is natural to regard these stations as nodes within the network. In contrast, dockless systems afford users the flexibility to pick up and drop off bikes anywhere, without the constraint of fixed stations. If we consider these sources and destinations to be a virtual station, this creates a graph with an unlimited number of stations, creating a complex network where nodes may potentially have very low degree scores as very few trips begin or end at precise locations. This requires a method to optimise the graph by reducing the number of nodes (virtual  stations) in the network. While our previous work \cite{CNCR2024} reassigned non-station source and destination nodes to their closest fixed station, the goal here is the optimised expansion of the network before this reassignment task, in order to retain as much spatial information regarding trips, as possible.

A novel strategy was developed to use hierarchical agglomerative clustering (HAC) to group geographically close (starting or ending trip) locations with logical thresholds defined to identify \emph{candidate} locations for new stations that consider \emph{current} fixed stations. HAC is a bottom-up clustering algorithm that begins by considering each data point as an individual cluster and merges the clusters based on a distance measurement until a single cluster containing all of the data points emerges \cite{FENG201712422}. The distance metric used for geographic locations was the Haversine distance, shown in equation  \ref{eqn:hav_eq}. The function calculates the shortest distance between two points on a sphere and was chosen as it remains accurate for computations at small distances unlike calculations based on the spherical law of cosine \cite{Agramanisti_Azdy_2020}. 
  
    \begin{equation}
    \label{eqn:hav_eq}
    \footnotesize
    d=2R\arcsin\sqrt{\sin^2(\frac{\varphi_1-\varphi_2}2)
    +\cos\varphi_1\cos\varphi_2\sin^2(\frac{\lambda_1-\lambda_2}{2})} 
    \end{equation}

Equation \ref{eqn:hav_eq} shows the distance function where $R$ is the radius of Earth; $\varphi_1$ and $\lambda_1$ represent latitude and longitude for Location $A$; $\varphi_2$ and $\lambda_2$ represent latitude and longitude for Location B; and $d$ is the distance between two locations. The Complete Linkage criterion was used to determine the distance between two clusters based on the largest distance over all possible pairs \cite{RAMOSEMMENDORFER2021106990}.    

\vspace{2mm}
\textbf{Preprocessing.} Pre-existing fixed stations were set as immovable locations and set as their own group’s centroid. To adhere to the criterion of groups’ centroids being at least 50 metres apart, any location that was within a 50-metre radius of a fixed station was assigned to that station’s group and was excluded from clustering.

\subsection{Station Ranking and Selection}
\label{subSec:station_selection}

After conducting several analyses of the data, particularly focusing on instances where a high number of distinct locations were less than three meters apart, we established a set of parameter settings before initiating the clustering process. These settings formed a set of rules that governed the output of the clustering algorithm.


\begin{rule2}
Cluster-Boundary. \\
The distance between 2 locations $L_1$ and $L_2$ inside a given cluster $C$ may not exceed 100m.
\end{rule2}

\begin{rule2}
Cluster-Proximity. \\
The distance between any pair of cluster centroids $C_{ci}$ and $C_{cj}$ cannot be less than 50m.
\end{rule2}

\begin{rule2}
Degree-Threshold. \\
The degree $D$ for candidate station $S$ cannot be less than $S_{min}$, the minimum degree from the set of fixed stations.
\end{rule2}

\begin{rule2}
Secondary-Distance. \\
A second threshold was set to ensure that nodes which represent a cluster are not within 250m of a station.
\end{rule2}

The procedure for selecting stations from candidate stations is outlined below:
   
\begin{enumerate}
     \item Non-fixed station locations underwent clustering. The trips starting and ending at each location in each cluster. 
     
    \item As rule conflicts will occur if all candidates become fixed stations, clusters are  ranked in descending order for degree metrics. Before each candidate station is  converted to a fixed station, it is checked against each of the rules. Unless all 4 rules are true, the candidate station is rejected.   
    
    \item When the process completes, unconverted candidate locations are reassigned to the nearest station.
\end{enumerate}

\subsection{Community Detection}

In order to understand the behaviour of new stations, a community detection algorithm was used to study the spatiotemporal patterns at different levels of temporal granularity. Community detection was chosen as it can identify sub-regions, in which the nodes are more strongly connected to one another than to nodes in other sub-regions. Thus, community detection was used to reduce the complexity of the network and to facilitate the study of the network’s underlying structure as new stations (nodes) are added. Furthermore, it can serve as a validation of the overall network through the verification of good partitions \cite{SC21}. To ensure a robust validation, community detection is performed at three different levels of temporal granularity across three different network structures.

\vspace{2mm}
\noindent \textbf{Temporal Granularity.}
\begin{itemize}
    \item $T_{Null}$: includes no temporal features.
    \item $T_{Day}$: uses the day of the week that trips took place.
    \item $T_{Hour}$: uses the time of day that trips began.
\end{itemize}

The objective of experimenting with different levels of temporal granularity using the inherently spatial data was to detect clusters of stations that exchange many trips and that have similar temporal and/or spatial characteristics. Thus, providing a better understanding of usage patterns and a valuable decision-support tool for fleet management. 

The Louvain algorithm is a popular greedy community detection algorithm. This algorithm was chosen for community detection due to its rapid convergence properties, high modularity, hierarchical partitioning and its ability to incorporate weighted edges \cite{Xinyu.2015}. The Luovain algorithm presented in Algorithm \ref{alg:station-selection} was run on three slightly different bidirectional graphs.

\begin{algorithm}[H]
\caption{Station Selection Algorithm}
\begin{algorithmic}[1]
\Require List of new station candidates; List of pre-existing stations
\Ensure Selected new stations
\State Threshold $\gets$ minimum degree of pre-existing stations
\For{each station candidate}
    \State score of the station $\gets$ degree
    \If{degree $<$ Threshold}
        \State score of the station $\gets$ 0
    \ElsIf{distance to the nearest pre-existing station $\leq$ 0.25}
        \State score of the station $\gets$ 0
    \EndIf
\EndFor

\Repeat
    \For{each pair of station candidates with score $>$ 0}
        \If{distance between the two stations $\leq$ 0.25}
            \State score of the lower-degree station $\gets$ 0
        \EndIf
    \EndFor
\Until{no station candidates with score $> 0$ are near each other}
\State Sort station candidates based on the score \\
\Return station candidates with score $> 0$ as new stations
\end{algorithmic}
\label{alg:station-selection}
\end{algorithm}

\vspace{2mm}
\noindent \textbf{Network Structures.}

Based on 3 temporal granularities $T_{Null}$, $T_{Day}$ and $T_{Hour}$, we have 3 structured graphs as follows:

\begin{itemize}
    \item  $G_{Basic}$ was constructed with stations represented by nodes, and trips represented by edges, weighted by the number of trips.
    \item  $G_{Day}$ has a similar node structure but each trip is now represented by a unique edge with a property denoting the day of the week that the trip took place.
    \item  $G_{Hour}$ is similar to $G_{Day}$ but each edge contains a property indicating the time of the day that the trip started. 
\end{itemize}

Modularity is the objective function used by the Louvain algorithm. It was also used to evaluate the resulting community structures. Modularity quantifies the quality of an assignment of nodes to a community by comparing the number of edges that are within communities to the number of edges expected if the edges were placed at random \cite{mod}. Modularity scores are scaled between -1 and +1, with a positive score indicating the presence of a community structure.  
 
 \begin{equation}
 \label{eqn:mod_eq}
 Q = \sum_{c_i\in C}[{\frac{\sum_{in}^{c_i}}{2m}-\frac{(\sum_{tot}^{c_i})^2}{4m^2}}]
 \end{equation}

The calculation for modularity (denoted by Q) is shown in equation \ref{eqn:mod_eq}, where $C$ is the set of communities; $m$ is the ratio of edges within the community to the total number of edges in the graph; $\Sigma^{c_i}_{in}$ is the sum of edges belonging to community c$_i$ over all of the nodes in the community; and $\Sigma^{c_i}_{tot}$ is the sum of the degree of all nodes in community c$_i$.

%
%

\section{Results and Discussion}

\subsection{Candidate Stations}

\begin{figure}[H]
\centering
\includegraphics[width=9cm]{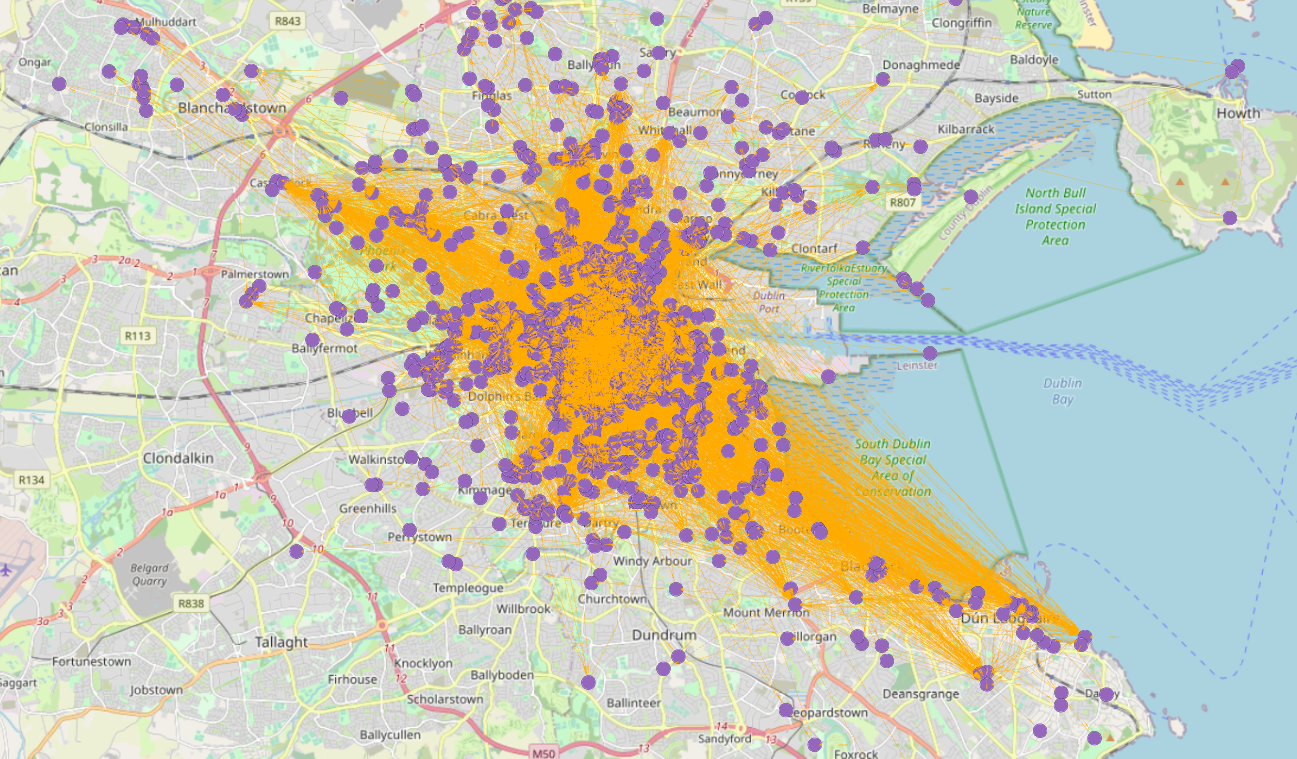}
\caption{The candidate graph generated by HAC, including the pre-existing stations. Nodes are shown in purple and edges in yellow.}
\label{fig:generated_data}
\end{figure}

Upon completing the graph construction outlined in Section \ref{subSec:GraphConstruction}, 1,172 clusters are generated, yielding a total of 1,080 potential locations for new stations, in addition to the existing 92 stations. The candidate graph, incorporating candidate stations, was aggregated and represented by weighted edges. The resulting graph is illustrated in Figure \ref{fig:generated_data}, and the corresponding data is summarized in Table \ref{tab:generated_data}.

\vspace{-5mm}
\begin{table}[H]
\centering
\caption{The details of the candidate graph generated by HAC}
\begin{tabular}{|l|r|}
\hline
\textbf{Measures} & \textbf{Numbers}  \\ \hline
\#nodes & 1,172  \\ \hline
\#undirected edges            & 8,240  \\ \hline
\#undirected edges (no loops) & 7,820  \\ \hline
\#directed edges             & 16,042 \\ \hline
\#directed edges (no loops)  & 15,604 \\ \hline
\#trips & 61,872 \\ \hline
\end{tabular}
\label{tab:generated_data}
\end{table}

\subsection{Selected New Stations}

\begin{figure}[H]
\centering
\includegraphics[width=9cm]{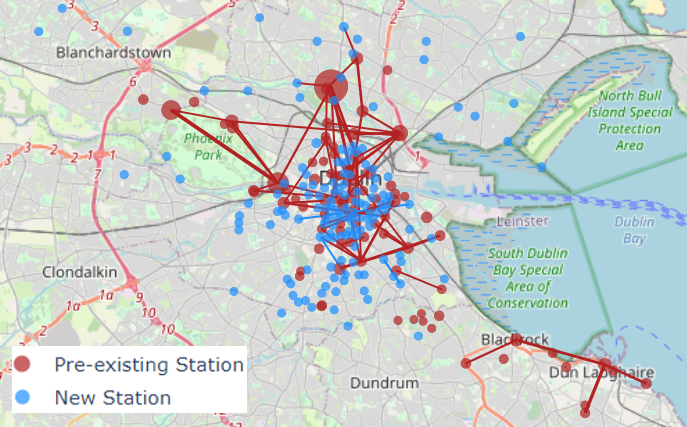}
\caption{The selected graph includes pre-existing stations and selected new stations. Node size is scaled according to the number of self-contained trips (self-edges). Edge width is scaled according to the number of directed trips between nodes (out-edges). Only edges with a weight in the top 1\% percentile of weights are shown.} 
\label{fig_new_stations}
\end{figure}

Following the execution of our station ranking and selection algorithm in Section \ref{subSec:station_selection}, a selected graph is generated, introducing 146 new stations and raising the overall count to 238 stations. The augmented graph is visualized in Figure~\ref{fig_new_stations} and detailed in Table \ref{tab:expanded-graph}. All trips from non-selected stations were redirected to the nearest existing station, ensuring that the total number of trips remains unchanged. Notably, the newly added stations are predominantly concentrated around Dublin City Centre, extending into the adjacent suburbs beyond the positions of the existing stations. 

\vspace{-4mm}
\begin{table}[h!]
\centering

\caption{Details of the selected Graph}
\begin{tabular}{|l|r|cc|cc|}
\hline
\multirow{2}{*}{}     & \multicolumn{1}{c|}{\multirow{2}{*}{Stations}} & \multicolumn{2}{c|}{Trips}     & \multicolumn{2}{c|}{Edges}    \\ \cline{3-6} 
& \multicolumn{1}{l|}{}    & \multicolumn{1}{c|}{From}  & \multicolumn{1}{c|}{To} & \multicolumn{1}{c|}{From} & \multicolumn{1}{c|}{To} \\ \hline
Pre-existing & 92 & \multicolumn{1}{r|}{54,670} & 54,727                   & \multicolumn{1}{r|}{6,437} & 6,310                    \\ \hline
Selected         & 146& \multicolumn{1}{r|}{7,202}  & 7,145                    & \multicolumn{1}{r|}{2,072} & 2,199                    \\ \hline
Total                 & 238& \multicolumn{2}{c|}{61,872}                 & \multicolumn{2}{c|}{8,509}                 \\ \hline
\end{tabular}
\label{tab:expanded-graph}
\end{table}

\subsection{Community Detection}
We need to be able to distinguish between new and existing stations. Are they spread across all communities? Can we update the table to reflect the location of these stations?

\subsubsection{Trips}

\begin{figure}[H]
\centering
    \includegraphics[width=9cm]{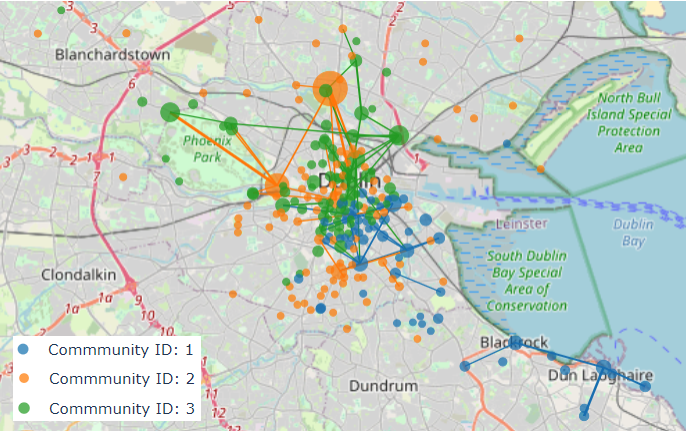}
    \caption{Community detection for $G_{Basic}$. Stations are coloured according to their respective community assignment.} 
    \label{figBasic}
\end{figure}

\vspace{-3mm}    
\begin{table}[h!]
    \caption{Detailed information about communities in $G_{Basic}$}
    \label{comms}
    \centering
    \scriptsize
    \begin{tabular}{|cl|ccc|cccc|}
    \hline
    \multicolumn{2}{|c|}{Community}   & \multicolumn{3}{c|}{Stations}        & \multicolumn{4}{c|}{Trips}   \\ \hline
    \multicolumn{1}{|c|}{ID} & Color  & \multicolumn{1}{c|}{Old} & \multicolumn{1}{c|}{New} & \multicolumn{1}{c|}{Total} & \multicolumn{1}{c|}{Within} & \multicolumn{1}{c}{Out}  & \multicolumn{1}{c|}{In}   & \multicolumn{1}{c|}{Total} \\ \hline
    \multicolumn{1}{|c|}{1}  & Blue   & \multicolumn{1}{r|}{40}  & \multicolumn{1}{r|}{18}  & 58  & \multicolumn{1}{r|}{12,012}  & \multicolumn{1}{r|}{5,238} & \multicolumn{1}{r|}{5,255} & 22,505   \\ \hline
    \multicolumn{1}{|c|}{2}  & Orange & \multicolumn{1}{r|}{4}   & \multicolumn{1}{r|}{94}  & 98  & \multicolumn{1}{r|}{9,158}   & \multicolumn{1}{r|}{4,078} & \multicolumn{1}{r|}{3,995} & 17,231  \\ \hline
    \multicolumn{1}{|c|}{3}  & Green  & \multicolumn{1}{r|}{48}  & \multicolumn{1}{r|}{34}  & 82  & \multicolumn{1}{r|}{24,494}  & \multicolumn{1}{r|}{6,892} & \multicolumn{1}{r|}{6,958} & 38,344  \\ \hline
\end{tabular}
\label{tab:T-Null-Network}
\end{table}

The first round of community detection used the $G_{Basic}$ graph with no temporal information.
The Louvain algorithm yielded three communities with a modularity score of 0.25. Given the low number of communities detected, this modularity score indicates that the sub-networks are non-trivial. The resulting communities are displayed in Figure~\ref{figBasic}. These communities have unique spatial properties. Stations belonging to the blue community are exclusively on the southside of Dublin; stations belonging to the orange community are typically less central than the other communities' stations and are spread around the suburbs of Dublin; while the green community's stations are generally in the city centre or are on the northside of Dublin.

The results of this community detection provided insight into the interactions between different communities, with a summary of the number of trips for each community provided in Table~\ref{comms}. For each community, this table details the exact number of trips that started and ended in that community (within), the number of trips that started in the community but ended in another (out), and the number of trips that started in another community but ended in the community (in). Approximately 74\% of the trips start and end in the same community. This finding is consistent with previous studies into the self-contained nature of BSSs in London and Beijing where 75\% and 77\% of the trips started and ended in the same communities \cite{f1,Lin_He_Peeta_2018}.
    
Around 50\% of all trips in the network start in the green community which is unsurprising as it is the most centrally located community. This community also has the highest proportion of within-community trips whereas the other 2 communities have a similar lower proportion of within-community trips. This could suggest trips involving stations in these 2 communities cover longer distances or are used for special purposes.
    
\subsubsection{Trips and Day of Week}
\begin{figure}[H]
    {\centering
    \includegraphics[width=9cm]{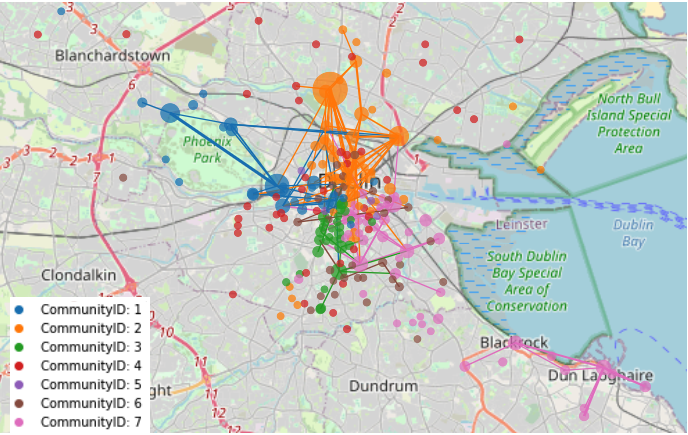}
    \caption{Community detection for $G_{Day}$.} 
    \label{figDay1}\par }
\end{figure}

\begin{figure}[H]
    \centering
    \includegraphics[scale=0.34]{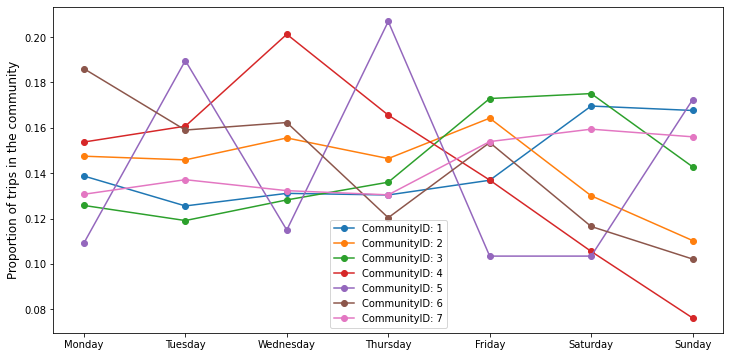}
    \caption{Daily travel patterns per community in $G_{Day}$} 
    \label{figDay2}
\end{figure}

The second round of community detection used the $G_{Day}$ network, incorporating the day of the week that a trip took place. On this occasion, 7 community structures were identified, with a modularity score of 0.32. The communities are displayed in Figure~\ref{figDay1} and Table \ref{tab:G-Day}, while the proportion of each community's trips that took place on each day of the week is shown in Figure~\ref{figDay2}.

Distinct temporal patterns can be seen among the larger communities. For instance, usage was lowest during the weekends in Community 2, Community 4, and Community 6. Bikes in these communities are likely largely used for weekday commuting purposes. These uncovered patterns could be used to assist with fleet re-balancing strategies. Conversely, usage peaked on Saturday in Community 1, Community 3, and Community 7. This could be explained by these communities' geographical locations and their stations’ proximity to popular weekend spots. Community 1 is mainly composed of stations that are within close proximity to the Phoenix Park and many stations in Community 7 are within a short distance of Blackrock/Dún Laoghaire. These uncovered patterns could be used to assist with fleet re-balancing strategies. For example, bikes could be moved from Communities 2, 4, and 6 to Communities 1, 3, and 7 each Friday night to prepare for the shift in demand over the weekend.

\vspace{-3mm}  
\begin{table}[H]
\centering
\caption{Detailed information about communities in $G_{Day}$}
\scriptsize
\begin{tabular}{|cl|rrr|rrrr|}
\hline
\multicolumn{2}{|c|}{Community}   & \multicolumn{3}{c|}{Station}         & \multicolumn{4}{c|}{Trips}                     \\ \hline
\multicolumn{1}{|c|}{ID} & Color  & \multicolumn{1}{c|}{Old} & \multicolumn{1}{c}{New} & \multicolumn{1}{c|}{Total} & \multicolumn{1}{c|}{Within} & \multicolumn{1}{c|}{Out}  & \multicolumn{1}{c|}{In}   & \multicolumn{1}{c|}{Total} \\ \hline
\multicolumn{1}{|c|}{1}  & Blue   & \multicolumn{1}{r|}{15}  & \multicolumn{1}{r|}{16}  & 31   & \multicolumn{1}{r|}{8,517}   & \multicolumn{1}{r|}{3,516} & \multicolumn{1}{r|}{3,522} & 15,555\\ \hline
\multicolumn{1}{|c|}{2}  & Orange & \multicolumn{1}{r|}{0}   & \multicolumn{1}{r|}{22}  & 22   & \multicolumn{1}{r|}{551}    & \multicolumn{1}{r|}{227}  & \multicolumn{1}{r|}{238}  & 1,016 \\ \hline
\multicolumn{1}{|c|}{3}  & Green  & \multicolumn{1}{r|}{14}  & \multicolumn{1}{r|}{16}  & 30   & \multicolumn{1}{r|}{3,983}   & \multicolumn{1}{r|}{3,995} & \multicolumn{1}{r|}{4,049} & 12,027\\ \hline
\multicolumn{1}{|c|}{4}  & Red    & \multicolumn{1}{r|}{0}   & \multicolumn{1}{r|}{27}  & 27   & \multicolumn{1}{r|}{551}    & \multicolumn{1}{r|}{179}  & \multicolumn{1}{r|}{170}  & 900  \\ \hline
\multicolumn{1}{|c|}{5}  & Purple & \multicolumn{1}{r|}{36}  & \multicolumn{1}{r|}{16}  & 52   & \multicolumn{1}{r|}{11,555}  & \multicolumn{1}{r|}{4,949} & \multicolumn{1}{r|}{4,933} & 21,437\\ \hline
\multicolumn{1}{|c|}{6}  & Brown  & \multicolumn{1}{r|}{0}   & \multicolumn{1}{r|}{32}  & 32   & \multicolumn{1}{r|}{1,411}   & \multicolumn{1}{r|}{450}  & \multicolumn{1}{r|}{414}  & 2,275 \\ \hline
\multicolumn{1}{|c|}{7}  & Pink   & \multicolumn{1}{r|}{27}  & \multicolumn{1}{r|}{17}  & 44   & \multicolumn{1}{r|}{16,328}  & \multicolumn{1}{r|}{5,660} & \multicolumn{1}{r|}{5,650} & 27,638\\ \hline
\end{tabular}
\label{tab:G-Day}
\end{table}

\subsubsection{Trips and Time of Day}

\begin{figure}[H]
{\centering
\includegraphics[width=9cm]{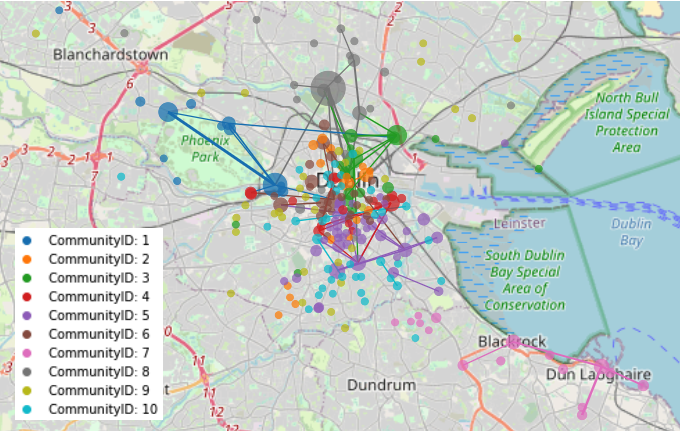}
\caption{Community detection for $G_{Hour}$ } \label{fig:hour1}\par }
\end{figure}

\begin{figure}[H]
\centering
\includegraphics[scale=0.34]{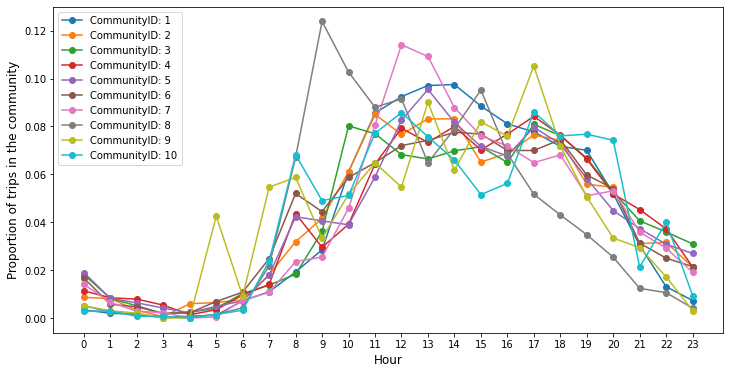}
\caption{Hourly Travel patterns per community in $G_{Hour}$.} \label{fig:hour2}
\end{figure}

The final round of community detection used the $G_{Hour}$ network, which utilised the time of the day that trips began, detecting 10 communities with a strong modularity score of 0.54. These communities provide further support that the network is heavily influenced by spatiotemporal patterns. The communities are displayed in Figure~\ref{fig:hour1} and Table \ref{tab:G-Hour}. While, a breakdown of usage per hour of the day is shown in Figure~\ref{fig:hour2}.

\begin{table}[H]
\centering
\caption{Detailed information about communities in $G_{Hour}$}
\scriptsize
\begin{tabular}{|cl|rrr|rrrr|}
\hline
\multicolumn{2}{|c|}{Community}   & \multicolumn{3}{c|}{Station}         & \multicolumn{4}{l|}{Trips}   \\ \hline
\multicolumn{1}{|c|}{ID} & Color  & \multicolumn{1}{c|}{Old} & \multicolumn{1}{c|}{New} & \multicolumn{1}{c|}{Total} & \multicolumn{1}{c|}{Within} & \multicolumn{1}{c|}{Out}  & \multicolumn{1}{c|}{In}   & \multicolumn{1}{c|}{Total} \\ \hline
\multicolumn{1}{|c|}{1}  & Blue   & \multicolumn{1}{r|}{9}   & \multicolumn{1}{r|}{4}   & 13   & \multicolumn{1}{r|}{5,422}   & \multicolumn{1}{r|}{1,706} & \multicolumn{1}{r|}{1,704} & 8,832 \\ \hline
\multicolumn{1}{|c|}{2}  & Orange & \multicolumn{1}{r|}{13}  & \multicolumn{1}{r|}{11}  & 24   & \multicolumn{1}{r|}{1,774}   & \multicolumn{1}{r|}{1,930} & \multicolumn{1}{r|}{1,944} & 5,648 \\ \hline
\multicolumn{1}{|c|}{3}  & Green  & \multicolumn{1}{r|}{11}  & \multicolumn{1}{r|}{9}   & 20   & \multicolumn{1}{r|}{4,762}   & \multicolumn{1}{r|}{4,062} & \multicolumn{1}{r|}{4,083} & 12,907\\ \hline
\multicolumn{1}{|c|}{4}  & Red    & \multicolumn{1}{r|}{10}  & \multicolumn{1}{r|}{9}   & 19   & \multicolumn{1}{r|}{2,379}   & \multicolumn{1}{r|}{2,833} & \multicolumn{1}{r|}{2,825} & 8,037 \\ \hline
\multicolumn{1}{|c|}{5}  & Purple & \multicolumn{1}{r|}{14}  & \multicolumn{1}{r|}{0}   & 14   & \multicolumn{1}{r|}{8,313}   & \multicolumn{1}{r|}{4,974} & \multicolumn{1}{r|}{4,991} & 18,278\\ \hline
\multicolumn{1}{|c|}{6}  & Brown  & \multicolumn{1}{r|}{15}  & \multicolumn{1}{r|}{14}  & 29   & \multicolumn{1}{r|}{3,234}   & \multicolumn{1}{r|}{3,613} & \multicolumn{1}{r|}{3,656} & 10,503\\ \hline
\multicolumn{1}{|c|}{7}  & Pink   & \multicolumn{1}{r|}{6}   & \multicolumn{1}{r|}{18}  & 24   & \multicolumn{1}{r|}{4,186}   & \multicolumn{1}{r|}{1,161} & \multicolumn{1}{r|}{1,175} & 6,522 \\ \hline
\multicolumn{1}{|c|}{8}  & Gray   & \multicolumn{1}{r|}{9}   & \multicolumn{1}{r|}{17}  & 26   & \multicolumn{1}{r|}{5,450}   & \multicolumn{1}{r|}{2,310} & \multicolumn{1}{r|}{2,256} & 10,016\\ \hline
\multicolumn{1}{|c|}{9}  & Olive  & \multicolumn{1}{r|}{1}   & \multicolumn{1}{r|}{30}  & 31   & \multicolumn{1}{r|}{767}    & \multicolumn{1}{r|}{221}  & \multicolumn{1}{r|}{207}  & 1,195 \\ \hline
\multicolumn{1}{|c|}{10} & Cyan   & \multicolumn{1}{r|}{4}   & \multicolumn{1}{r|}{34}  & 38   & \multicolumn{1}{r|}{1,912}   & \multicolumn{1}{r|}{863}  & \multicolumn{1}{r|}{832}  & 3,607 \\ \hline
\end{tabular}
\label{tab:G-Hour}
\end{table}

Communities that experience a spike in demand between 7 am and 9 am and again at around 5 pm, such as Community 9 and Community 10, are likely mainly used by commuters. These communities have a large proportion of stations in the suburbs and/or in the city centre. On the other hand, usage in Communities 1 and 7 spikes at around midday, and are comprised of stations around the Phoenix Park and Dún Laoghaire, respectively.

%
%
\newpage
\section{Conclusions}
This study utilised Moby Bike trip data to reveal temporal and spatial patterns within the complex BSS network. The geospatial nature of the data motivated the use of graph theory to accomplish this objective. This study was initially impeded by the sheer number of locations and software limitations. While research with similar objectives exists, there are a limited number of research papers available that examine the usage patterns of dockless BSSs. Therefore, a novel approach to intelligently condense the size of the network using hierarchical clustering and logical thresholds was developed. The results from this network optimisation process were validated visually and by comparing the number of new stations to the number of pre-existing stations. The output of the optimisation process could potentially be used by system operators to identify locations for new fixed stations. 

Community detection at three different levels of temporal granularity was performed on the optimised graph. The temporal dependence of the network is exposed by examining the community structures. Spatiotemporal fluctuations are observed in the detected communities using the day of the week and the time of day. Peak usage appears to be driven by commuting hours and specific leisure usage. Furthermore, the largely self-contained nature of the resulting communities, revealed by their strong modularity scores, suggests that bikes are not freely flowing between communities, which emphasises the importance of redistributing the bikes. These findings enable focused policies to optimise fleet rebalancing strategies and to meet user demand.

The outcomes of this research enables evidence-based policies to address network expansion and supply shortages. However, this study has several limitations. The criterion for clustering, that no two locations in a single cluster should exceed 100 metres, and the threshold that a new station must be at least 250 meters away from all other stations, were not motivated by empirical evidence. Instead, these distances were set based on pragmatic reasoning. Another limitation of this study is that it does not explore external factors such as weather or urban amenities that influence BSS usage. This study suggests an interconnection between leisure spots and bike-sharing usage patterns, which implies that a more holistic approach would be useful in uncovering additional network dependencies. Finally, this work did not experiment with different community detection algorithms. 

Future studies should compare the results of a range of community detection algorithms, such as the Infomap algorithm and the Label Propagation algorithm. Further research should also investigate the effect of different graph optimisation strategies on community detection, particularly if more computational resources are available to allow for larger graphs.

\vspace{5mm}
\noindent
\textbf{Funding Acknowledgement.} This work was supported by Science Foundation Ireland through the Insight Centre for Data Analytics (SFI/12/RC/2289\_P2) and the Vistamilk SFI Research Centre (SFI/16/RC/3835).

%
%
\bibliographystyle{IEEEtran}
\bibliography{main}

\end{document}